# UAV Communications: Impact of Obstacles on Channel Characteristics

Kamal Shayegan
Telecommunication Company of Iran
K1470sh@gmail.com

***Abstract*— In recent years, Unmanned Aerial Vehicles (UAVs) have been utilized as effective platforms for carrying Wi-Fi Access Points (APs) and cellular Base Stations (BSs), enabling low-cost, agile, and flexible wireless networks with high Quality of Service (QoS). The next generation of wireless communications will rely on increasingly higher frequencies, which are easily obstructed by obstacles. One of the most critical concepts yet to be fully addressed is positioning the UAV at optimal coordinates while accounting for obstacles. To ensure a line of sight (LoS) between UAVs and user equipment (UE), improve QoS, and establish reliable wireless links with maximum coverage, obstacles must be integrated into the proposed placement algorithms. This paper introduces a simulation-based measurement approach for characterizing an air-to-ground (AG) channel in a simple scenario. By considering obstacles, we present a novel perspective on channel characterization. The results, in terms of throughput, packet delivery, packet loss, and delay, are compared using the proposed positioning approach.**

## I. Introduction

In recent years, Unmanned Aerial Vehicles (UAVs) have gained popularity across various applications [1]. Examples include surveillance and reconnaissance, firefighting and rescue operations, remote sensing and exploration, pesticide spraying, geophysical surveys, logistics, and payload transport [2]. Beyond these general applications, providing wireless broadband connectivity is one of the most recent uses of UAVs. In scenarios where network infrastructure is absent or necessary to increase network capacity to meet users' traffic demands, UAVs offer an effective solution for maximizing user connectivity.

However, flying networks present significant challenges. Firstly, radio link disruptions may occur due to the high mobility of UAVs [4]. Additionally, there is inter-flow interference between neighboring UAVs, which must remain close to one another to establish high-capacity air-to-air radio links. These challenges were addressed in [5], [6]. For UAV networks to be effective, they must provide low latency, high throughput, and maximum coverage tailored to users' traffic demands [7]. Positioning UAVs efficiently is therefore essential. In other words, deploying a UAV to a location without traffic demand is counterproductive, especially when users in other areas are awaiting service. Consequently, user demand should guide the deployment of UAVs to ensure optimal performance.

When it comes to the UAV placement problem, the state of the art has only focused on UAVs acting as APs and BSs, aiming at covering the ground users and improving the Quality of Service (QoS) [8], [9]. In other words, most relevant works in placement algorithms rely on deploying the UAVs in an optimal position without considering the environmental conditions, namely obstacles. [10]–[12]. UAVs are responsible for providing communication services to users within their transmission range, the position of the UAV determines how much capacity each user gets [13]. UAV repositioning due to blockage of LoS with ground users, which in turn happens because of obstacles, is a subject that has not been investigated so far. In this paper, we simulate a simple scenario of wireless communication, including a UAV acting as a Wi-Fi AP, and a user aiming to provide a broadband communication link. We use the ns-3 to simulate the scenario and measure the channel characteristics. The simulation will be advanced by considering the UAV in a LoS situation with UE, and then in another scenario, the non-LoS position will be investigated. The derived throughput, packet delivery/loss, and delay will be compared together.

This paper's main contribution is to present a simulation measurement approach for characterizing an AG channel in a simple scenario for airborne communication by UAV, as well as to investigate the impact of the obstacle on QoS parameters and channel characteristics. For this purpose, considering the positioning approach, After putting an obstacle in the scenario between the user and the UAV, we characterize the channel parameter and articulate the effects of blocking LoS on the link quality. After changing the position of the UAV to a set of coordinates where the building is no longer an obstacle, the LoS must be re-established and the QoS must be increased. The performance of the network is evaluated through simulation using ns-3. Our measurements have been focused on throughput, Packet Delivery Ratio (PDR), and delays. The achieved results will be compared.

The rest of the paper is organized as follows. Section II presents the state of the art and related work. Section III defines the system model, including the network architecture. Section IV addresses the measurement setup, including the simulation setup, the measurement metrics, and simulation results. Section V discusses the simulation results. Finally, Section VI points out the main conclusions and directions for future work.

## II. State of the Art

Various measuring techniques and strategies are described as the state of the art in the context of AG communication channel characterization by UAVs. In terms of positioning, the related works can be divided into three categories based on the impact of obstacles on providing wireless network access by UAVs. In what follows, we refer to the works related to each of these categories.

*1) Positioning of UAVs in obstacle-free environments.* The research works in this category propose solutions to optimize the positions of UAVs for providing low-latency and high-throughput communications to ground users. However, they just work in obstacle-free environments. In [14], the authors analyze the coverage and rate performance of UAV-based wireless communication in the presence of underlaid D2D communication links. In [15], the authors propose the optimal 3D deployment of multiple UAVs to maximize the downlink coverage performance while using a minimum transmission power. In [11], the focus is on using UAVs as mobile-based stations while minimizing the number of UAVs employed to cover the users. In [15], the authors propose an algorithm to optimize multiuser scheduling and association jointly with the UAVs' trajectory and power control.

*2) Blocked LoS by dynamic obstacles for stationary BSs.* The research works in the this category address the probability of users being blocked by dynamic obstacles, while the provider is a stationary BS. In [16], the authors propose an end-to-end system of infrastructure-mounted LiDAR sensors to capture the dynamic obstacles. In [17], a solution that proactively predicts dynamic link blockages by using a camera is proposed. In [18], It demonstrates how computer vision enables look-ahead prediction in a millimeter-wave channel blockage scenario before the blockage actually occurs. However, the works in this category did not take advantage of UAVs to provide a wireless network.

*3) using a 3D pre-defined environment map to calculate the effect of obstacles on shadowing and LoS.* The research works in this category focus on positioning the UAVs based on a pre-defined 3D map of the area that has already been uploaded [22]. They consider the reflection effects of obstacles to characterize the wireless Channel in terms of path loss and shadowing. In [19], The authors proposed a resource allocation algorithm for millimetres-wave multicast systems that use multiple UAVs with Intelligent Reflecting Surfaces (IRS). In [20], an optimization problem is formulated to maximize the minimum capacity among users by jointly optimizing the 3-D position and power allocation of the UAVs. In [21], the blockage effect caused by buildings is investigated according to the 3-D geographic information for providing LoS to the users. A tractable model of urban UAV networks, where the obstacle reflection can be used to recover the LoS communication when the direct path is blocked is proposed in [21]. This category only focuses on the possibility of passive beamforming and the effect of blockage on LoS and doesn't work on positioning solutions.

In terms of AG channel characterization, the majority of the work did not take into account the UAV positioning concept in non-obstacle-free scenarios to optimize the characterization of the channel. The authors of [15] presented a model for characterizing AG and ground-to-air (GA) channels. The UAV was hovering at different altitudes and, consequently, at different line-of-sight (LoS) distances from the UE. This work has analyzed the channel model in terms of path loss and fast-fading components. In [7], the authors computed the path loss exponent for AG networks while the UAV is flying over a campus area and an open field using Received Signal Strength (RSS) values. The UDP throughput of the air-ground-air lines is also measured by the authors.

## III. System Model

We consider an architecture to consist of a UAV acting as an Ap, a user that is waiting to be served by the UAV, and a BS that serves the UAV. Figure 1, shows that the UAV has a LoS with the user in both positions 1 and 2. However, there is no LoS with the user in Figure 2. when the building is established as an obstacle between the user and the UAV in position 2. Because of the height of BS, the UAV is served by BS even in the presence of buildings. Assume there is a LoS in the first step and the UAV is at a coordinate like position 1 in Figure 2. Following this measurement, in the second step, we simulate a non-LoS scenario by repositioning the UAV to a coordinate like position 2 in Figure 3. The outcomes are compared with the scenario in which we move the UAV from position 2 to position 1 when there is an obstacle.

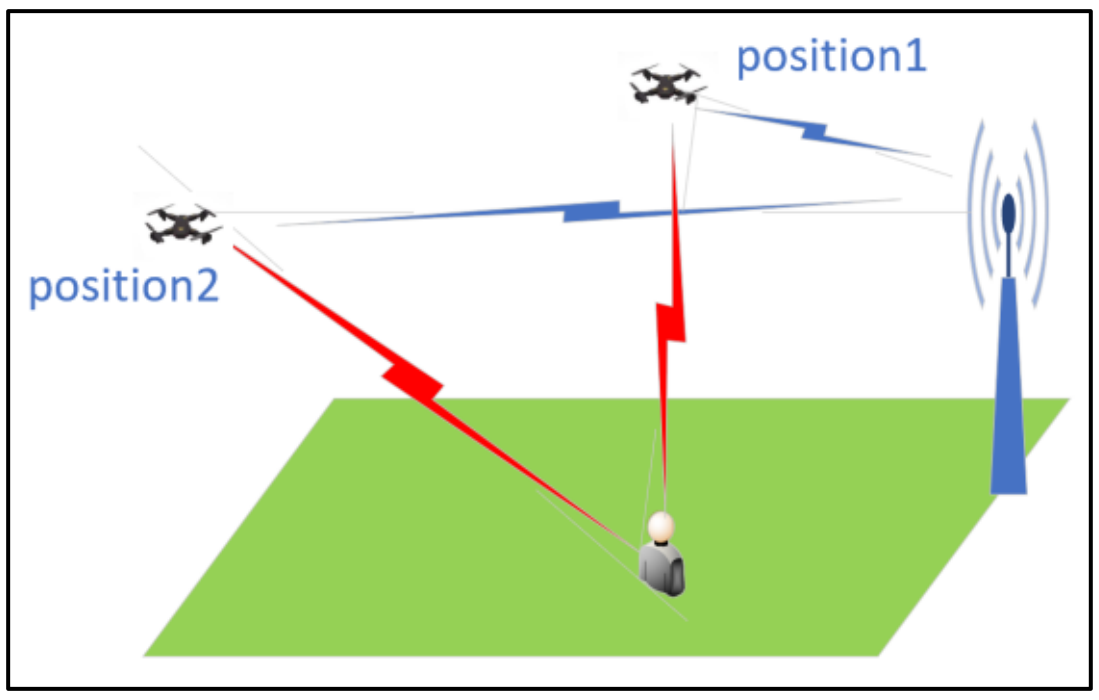

Fig. 1. Wireless network providing by UAV in a scenario without obstacle

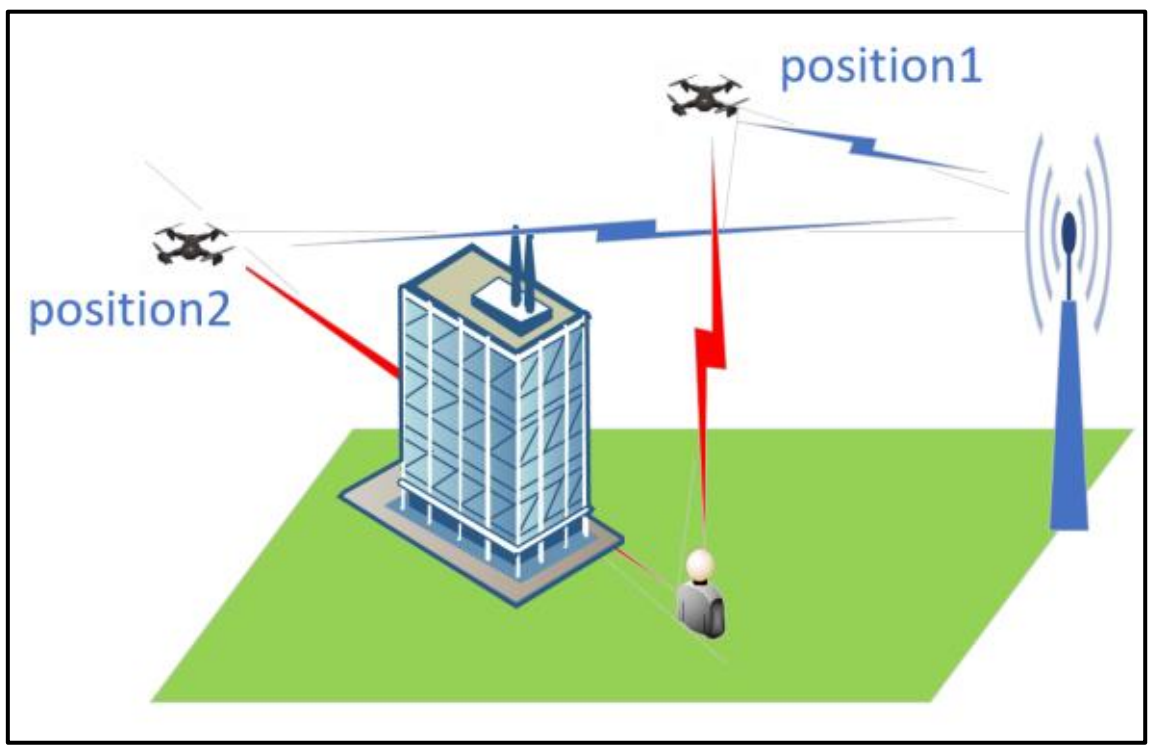

Fig. 2. Wireless network providing by UAV in a scenario with obstacle

## IV. Measurements Setup

The methodology followed for the Measurements from the simulation of both scenarios is presented in this section, including the simulation setup, the measurement metrics, and the results.

### *A. Simulation Setup*

As Fig. 4 shows, to evaluate the performance of the network and make measurements for the AG channel between the provider (UAV) and ground user (UE), we use the ns-3 simulator. In the first scenario, because the UAV should be stationary ns3::ConstantPositionMobility- Model is installed on the UAV acting as AP, and the coordinates (30.0, 00.0, 10.0) have been defined as the first UAV position, and a ground user, or UE, is supposed to be at (0.0, 0.0, 0.0). We have deployed the Wi-Fi AP at an altitude of 10 meters. A TCP receiver has been installed on the AP, and a TCP/UDP transmitter has been installed on the station. The Friis path loss model is employed for propagation loss model. A summary of the ns-3 simulation parameters considered is presented in Table I.

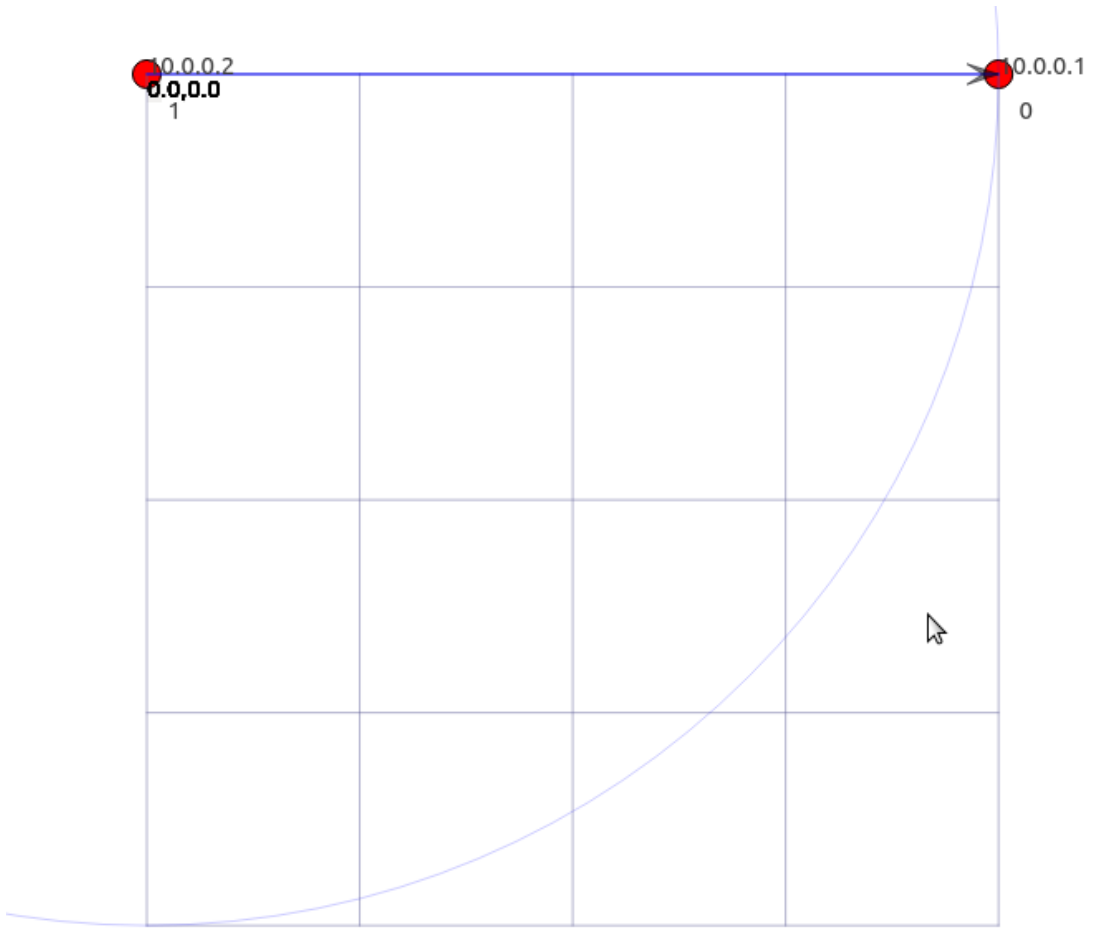

Fig. 3. Manifestations for the first scenario. A user in (0.0, 0.0, 0.0) coordinate and a UAV in (30.0, 00.0, 10.0).

In the second scenario, we have deployed a building as an obstacle between the AP and the ground user. In terms of mobility models, we have used ns3::ConstantPosition-MobilityModel for the Wi-Fi AP and the same for the user. Since the position of the obstacle in this scenario is constant, the MobilityBuildingInfoModel is used for building that is deployed as obstacles. As we can see in Fig. 3, the coordinates of the building where it is between AP and the user are x_min = 10.0; x_max = 20.0; y_min = 0.0; y_max = 50.0; z_min = -30.0; z_max = 30.0. In terms of the propagation loss model, we have applied HybridBuildings- PropagationLossModel in this scenario. A summary of the ns-3 simulation parameters considered in this scenario is presented in Table II.

TABLE I. SUMMARY OF THE OF THE NS-3 SIMULATION PARAMETERS CONSIDERED IN FIRST SCERARIO

| | |
|---|---|
| Simulation time | (10 s initialization +) 1 s |
| Wi-Fi standard | WIFI_STANDARD_80211ac |
| frequency | 5 GHZ |
| Channel Bandwidth | 100Mbps |
| Guard Interval | 100 ms |
| Transport layer payload size | 1472 bit |
| Propagation delay model | Constant speed |
| Propagation loss model | Friis path loss |
| Remote station manager | ConstantRateWifiManager |
| Wi-Fi mode | DLT_IEEE802_11_RADIO |
| Mobility model | ConstantPositionMobilityModel |
| ErrorRateModel | YansErrorRateModel |
| Traffic type | UDP/TCP |
| Packet size | 1024 bytes |

## B. Measurements Metrics

This work measures some channel characteristics in a simulation environment with a novel approach and considers obstacles as significant challenges in wireless communication to provide LoS. The latest release of ns-3 does not employ a direct model to deploy obstacles in a specific scenario. In this paper, we present this approach and conduct experimental measurements in a simulation environment using man-made obstacles (e.g., buildings and trees) to learn how it works for future work. During the performance evaluation of the scenario, we consider three measurement metrics:

- **Tthroughput –** The average number of bits received per second by the Wi-Fi AP.
- **Packet Delivery Ratio (PDR) –** The total number of packets received by the Wi-Fi AP divided by the total number of packets generated by the station, as measured at each station
- **End-to-end delay –** The time it takes packets to reach the Wi-Fi AP's application layer since they were generated is measured in seconds and includes queuing, transmission, and propagation delays.

Besides the ns-3, Flow Monitor is used to extracting the measurement parameters. Trace metric and Wireshark are other tools that we have used in this measurement.

## C. Simulation Results

The simulation results obtained for both scenarios are presented in this sub-section. The results were obtained after considering 10 simulation runs for each scenario. The average throughput in the first scenario in a frequency band ( F) equal to 5 GHz is 51.819 Mbit/s, the packet delivery ratio (PDR) is 99%, the packet loss ratio is 1%, and the end-to-end delay is 6.61881 ms. Fig. 5 shows the variation of throughput over time.

TABLE II. SUMMARY OF THE NS-3 SIMULATION PARAMETERS CONSIDERED IN SECOND SCERARIO

| | |
|---|---|
| Simulation time | (10 s initialization +) 1 s |
| Wi-Fi standard | WIFI_STANDARD_80211ac |
| frequency | 5 GHz |
| Channel bandwidth | 100Mbps |
| Guard Interval | 100 ms |
| Transport layer payload size | 1472 bit |
| Propagation delay model | Constant speed |
| Propagation loss model | HybridBuildingsPropagationLoss-Model |
| Remote station manager | ConstantRateWifiManager |
| Wi-Fi mode | DLT_IEEE802_11_RADIO |
| Mobility model | ConstantPositionMobilityModel and MobilityBuildingInfo |
| Error Rate Model | YansErrorRateModel |
| Traffic type | UDP/TCP |
| Packet size | 1024 bytes |

In the second scenario, where F = 5GHz, and we have deployed the building as an obstacle, the obtained throughput has decreased and the delay has increased. The average throughput in this case was 41.5377 Mbit/s, the PDR was 99%, the packet loss ratio was 1%, and the end-to-end delay was 10.5386 ms. Fig. 6 shows the variation of throughput over time.

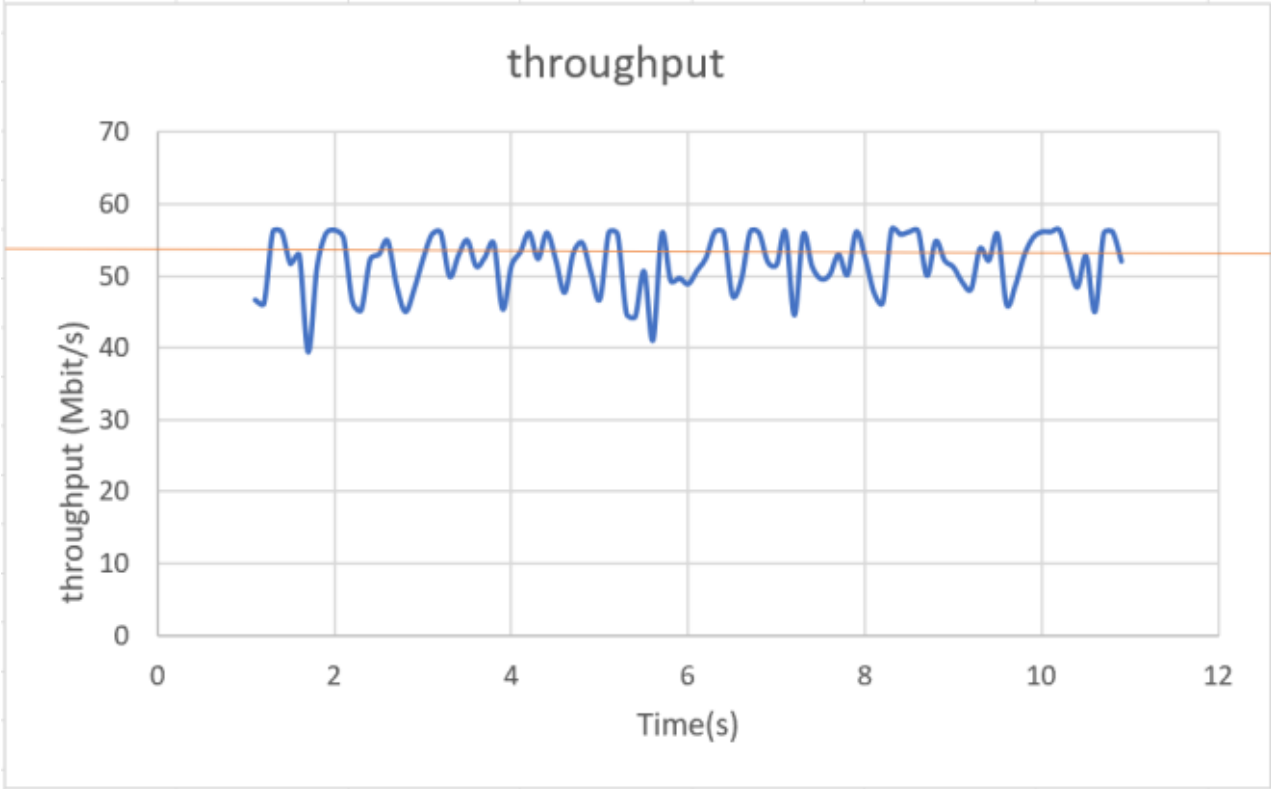

Fig. 4. Throghput variation for obstacle-free scenarion for F = 5 GHz

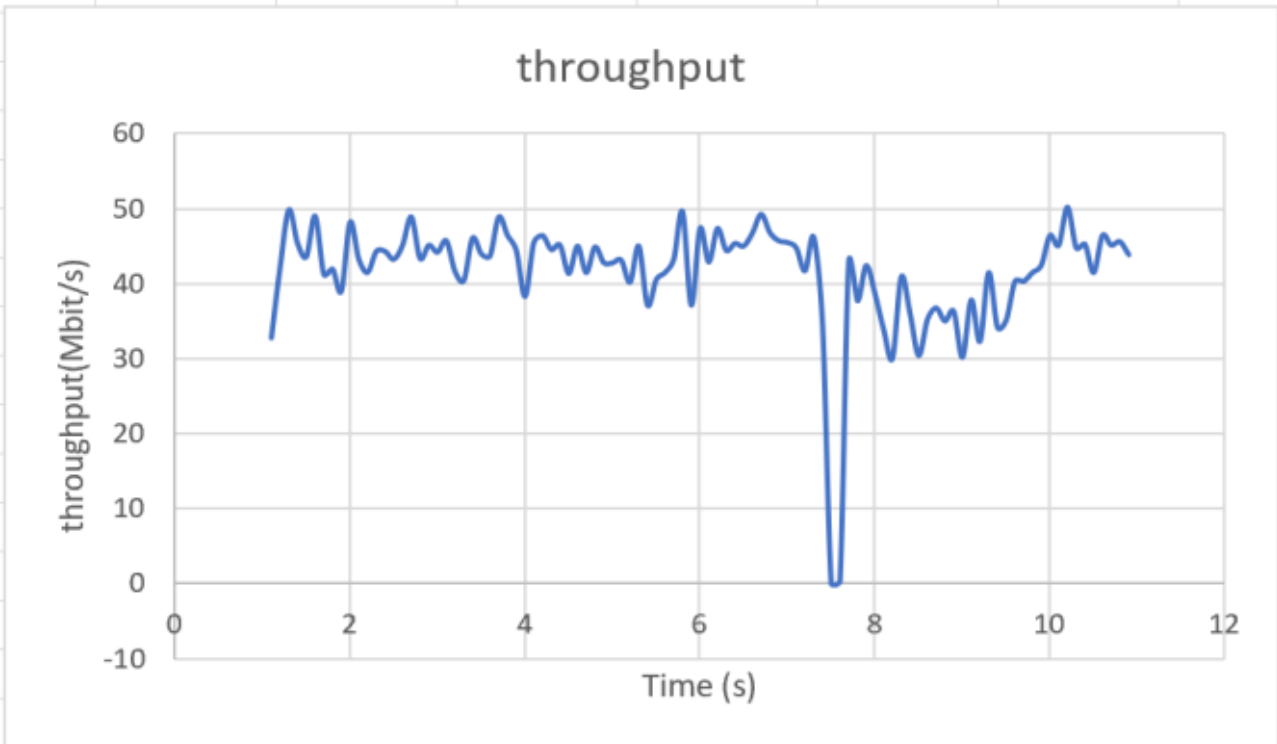

Fig. 5. Throghput variation for non-obstacle-free scenarion for F = 5GHz

In this case if the frequency increased to 10 GHz, with the same configuration, the throughput is decreased. The variation of 10 iterations of throughput have shown in in Fig. 7.

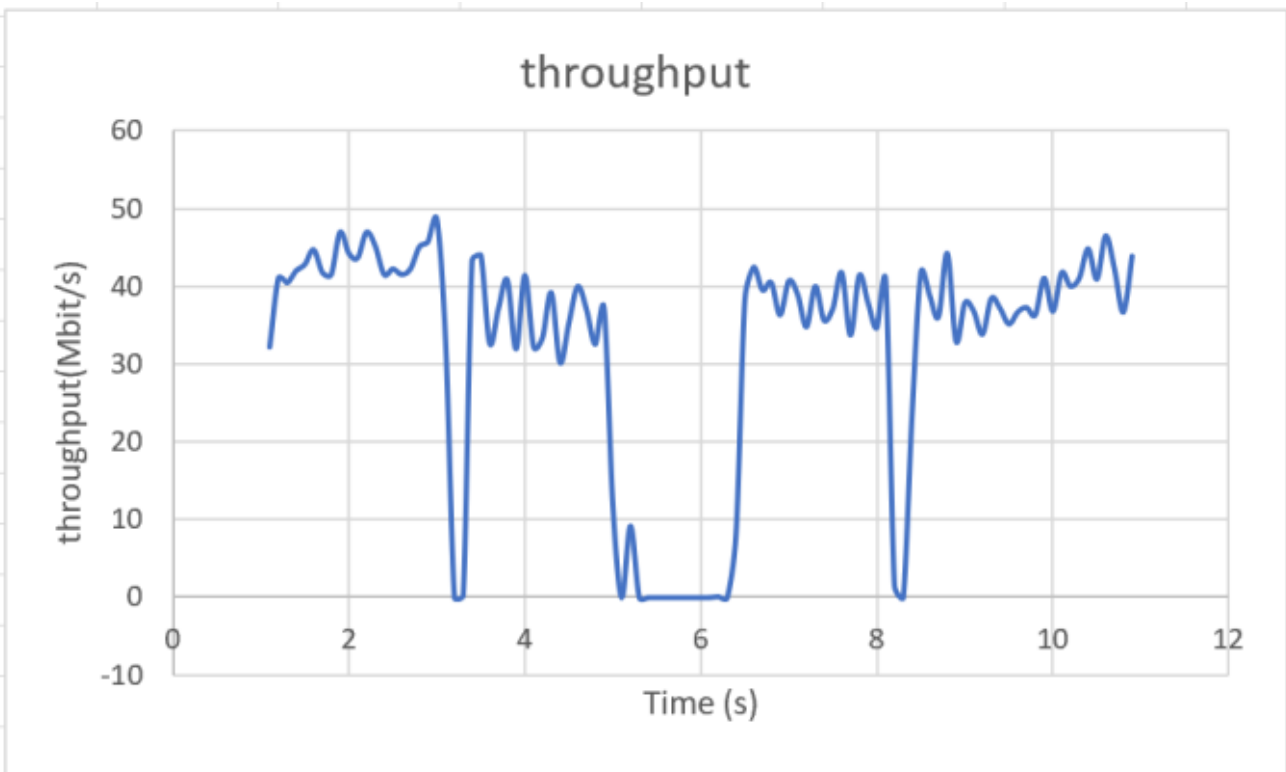

Fig. 6. Throghput variation for the second scenario for F = 10GHz

Finally, if we move the WI-FI AP from position 2 to position 1 (Fig. 2), where the building is no longer an obstacle, the LoS must be re-established, and the throughput and delay will return to their original levels. Fig. 8 shows the variation of throughput over time in position 2.

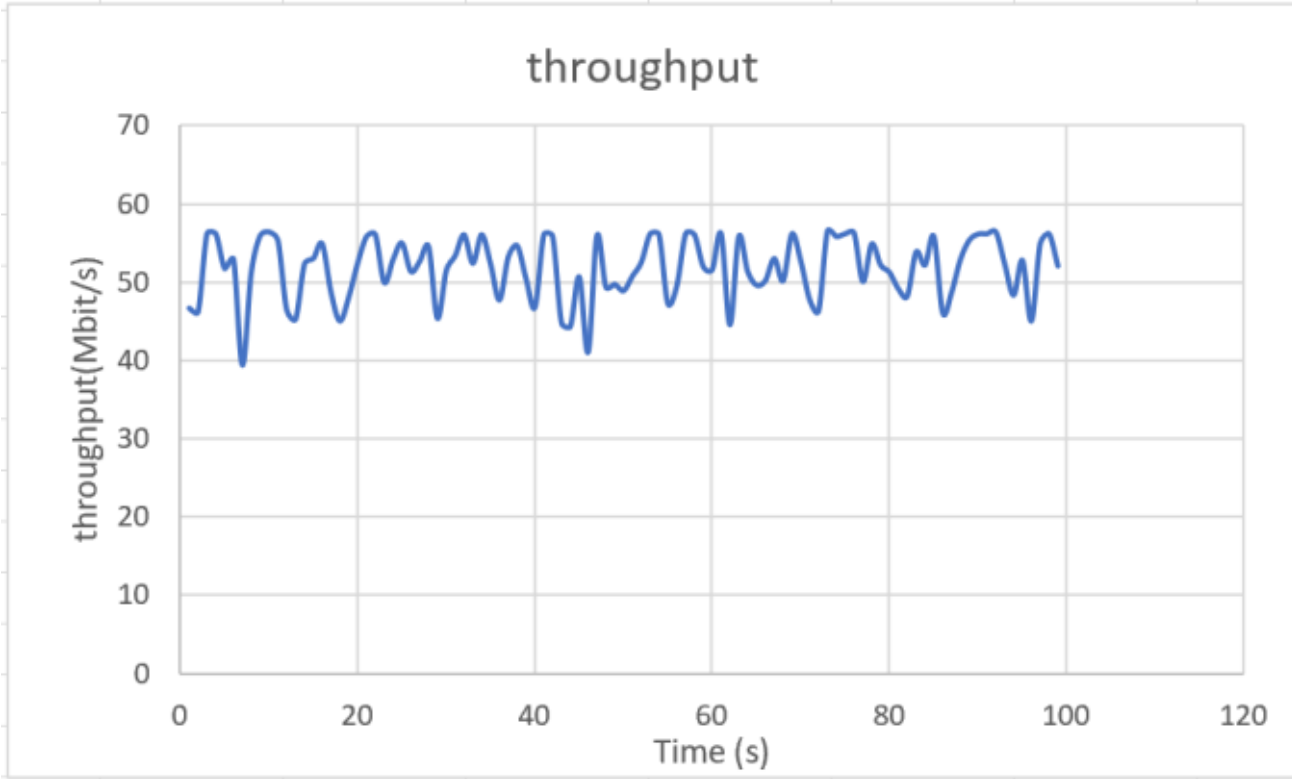

Fig. 7. Throghput variation for second scenarion in position2 and F = 5GHz

## V. Discussion

The optimal approach for conveying non-critical data, like video transmission, is to increase the bit rate while tolerating delays and errors. Sending vital data, such as control signals, calls for extremely high Quality of Service (QoS) standards in terms of error and delay, as well as low data rate requirements. LoS between AP and users affected by obstacles and providing the LoS by repositioning the UAV are the most important in terms of network efficiency. For instance, in an urban environment where there are a lot of obstacles, such as buildings and trees, we should consider LoS. The simulation results show that in the first scenario, the throughput is higher than in the second scenario. Due to the deployment of an obstacle in the middle of the scenario and increasing the acknowledgement time, the delay is increased in the second scenario. This is what we said at the beginning of the paragraph about the trade-off between data rate and delay.

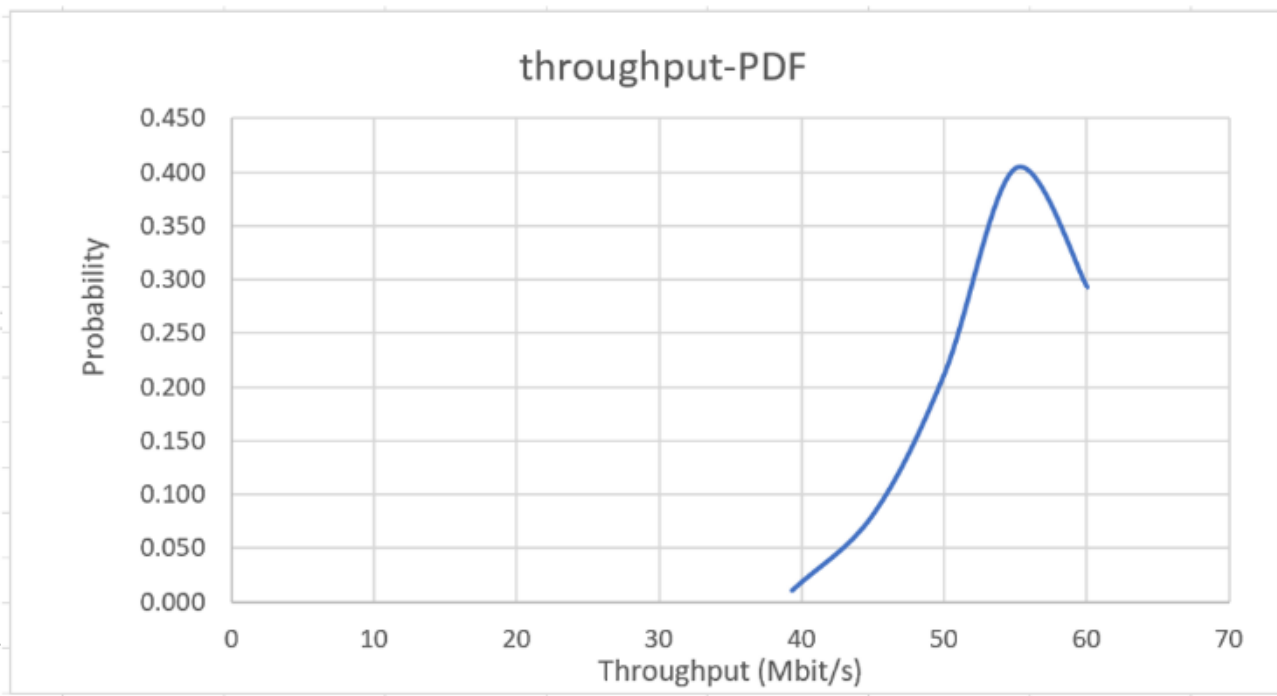

Fig. 8. probability distribution function of throughput after putting the obstacle in scenario for F = 5 GHz.

## VI. Main Conclusions

We presented a characterization of a wireless link between a UAV and a UE in this paper, where there is an obstacle between the provider and the ground user. We employed the Friis propagation loss model in an obstacle-free scenario and a hybrid building propagation loss model in case we had obstacles in the scenario. Considering the probability of LoS in an AG link, we measured the throughput, delay, and PDR of the channel. With this experiment, we have implicitly shown that the positioning of UAVs, taking into account obstacles and user demands, could help to improve performance and QoS, as well as provide maximum coverage for users. In Fig. 9 we can see the probability distribution function (PDF) of throughput after putting the obstacle in a scenario for F= 5 GHz.

For future works, considering a real urban area with more UAVs for covering wider areas can be a good research topic, as well as using signal processing solutions [23], optimizations [24], or Machine Learning [25].